\documentclass[aps,pra,showpacs,twocolumn,nofootinbib,10pt]{revtex4-1}
\usepackage{graphicx}
\usepackage{epstopdf}
\usepackage{amsmath}
\usepackage{amssymb}
\usepackage{mathrsfs}
\usepackage{amsthm}
\usepackage{bm}
\usepackage{url}
\usepackage{csquotes}
\MakeOuterQuote{"}

\newtheoremstyle{note}
  {\topsep/2}               % ABOVE SPACE
  {\topsep/2}               % BELOW SPACE
  {}                      % BODY FONT
  {\parindent}            % INDENT (empty value is the same as 0pt)
  {\itshape}              % HEAD FONT
  {.}                     % HEAD PUNCTUATION
  {5pt plus 1pt minus 1pt}% HEAD SPACE
  {}

\theoremstyle{note}

\newtheorem{theorem}{Theorem}
\newtheorem{lemma}{Lemma}

\newtheorem{remark}{Remark}

\def\vec#1{\bm{#1}} %% overriding the original command

\providecommand{\tr}{\operatorname{tr}}

\newcommand{\diag}{\operatorname{diag}}

\newcommand{\rep}{\mathrel{\widehat{=}}}

\newcommand{\rmA}{\mathrm{A}}
\newcommand{\rmB}{\mathrm{B}}

\newcommand{\rmE}{\mathrm{E}}

\newcommand{\rmM}{\mathrm{M}}
\newcommand{\rmS}{\mathrm{S}}
\newcommand{\rmT}{\mathrm{T}}

\newcommand{\spa}{\mathrm{Span}}

\def\eqref#1{\textup{(\ref{#1})}}  %% overriding the original command \eqref
\newcommand{\eref}[1]{Eq.~\textup{(\ref{#1})}}

\newcommand{\esref}[1]{Eqs.~\textup{(\ref{#1})}}

\newcommand{\fref}[1]{Fig.~\ref{#1}}

\newcommand{\sref}[1]{Sec.~\ref{#1}}
\newcommand{\Sref}[1]{Section~\ref{#1}}
\newcommand{\ssref}[1]{Secs.~\ref{#1}}

\newcommand{\thref}[1]{Theorem~\ref{#1}}

\newcommand{\Thsref}[1]{Theorems~\ref{#1}}

\newcommand{\lref}[1]{Lemma~\ref{#1}}

\newcommand{\cref}[1]{Conjecture~\ref{#1}}
\newcommand{\Cref}[1]{Conjecture~\ref{#1}}

\newcommand{\aref}[1]{Appendix~\ref{#1}}

\newcommand{\rcite}[1]{Ref.~\cite{#1}}

\begin{document}
\nocite{apsrev41Control}
\bibliographystyle{apsrev4-1}

\title{Einstein-Podolsky-Rosen correlations and Bell correlations in the simplest scenario}
\author{Quan Quan$^{1,2}$}
\author{Huangjun Zhu$^{3,4}$}
\email{hzhu1@uni-koeln.de, zhuhuangjun@gmail.com}
\author{Heng Fan$^{5,6,2}$}
\author{Wen-Li Yang $^{2,7}$}
\affiliation{$^1$State Key Laboratory of Low-Dimensional Quantum Physics and Department of Physics,
 Tsinghua University, Beijing 100084, China\\
$^2$Institute of Modern Physics, Northwest University, Xi'an 710069, China\\
$^3$Institute for Theoretical Physics, University of Cologne,  Cologne 50937, Germany\\
$^4$Perimeter Institute for Theoretical Physics,  Waterloo,  Ontario N2L 2Y5, Canada\\
$^5$Institute of Physics, Chinese Academy of Sciences, Beijing 100190, China\\
$^6$Collaborative Innovation Center of Quantum Matter, Beijing 100190, China\\
$^7$Shaanxi Key Laboratory for Theoretical Physics Frontiers, Xi'an 710069, China}

\date{\today}

\begin{abstract}
 Einstein-Podolsky-Rosen (EPR) steering is an intermediate type of quantum nonlocality which sits between entanglement and Bell nonlocality.
A set of  correlations is
Bell nonlocal if it does  not admit a local hidden variable (LHV) model,  while it is EPR nonlocal if it  does not admit a local hidden variable-local hidden state (LHV-LHS) model. It is interesting to know what  states can generate  EPR-nonlocal correlations  in the simplest nontrivial scenario, that is, two projective measurements for each party  sharing a two-qubit state. Here we show  that a two-qubit state can generate EPR-nonlocal full correlations (excluding marginal statistics) in this scenario if and only if it can generate Bell-nonlocal  correlations. If full statistics (including marginal statistics) is taken into account,  surprisingly, the same scenario can manifest  the simplest one-way steering and the strongest hierarchy between steering and Bell nonlocality. To illustrate these intriguing phenomena in simple setups,  several concrete examples are discussed in detail, which facilitates experimental demonstration.
 In the course of study, we introduce the concept of restricted LHS models and thereby derive a  necessary and sufficient semidefinite-programming criterion to determine the steerability of any bipartite state under given measurements. Analytical criteria are further derived in several   scenarios of strong theoretical and experimental interest.
\end{abstract}

\maketitle

\section{Introduction}
\label{Sintroduction}
Local measurements on entangled states can generate nonlocal correlations that cannot be reproduced by any classical mechanism \cite{EinsPR35, Schr35, Bell64,WiseJD07}. This  counterintuitive phenomenon is a  subject of continuous debate and inspiration for various new ideas \cite{HoroHHH09,Reid09,BrunCPS14}. It also plays a key role in many quantum information processing protocols, especially quantum key distribution (QKD) and  secure quantum communication \cite{Eker91,GisiRTZ02,BranCWS12}. The strongest nonlocal correlations can be detected by violation of Bell inequalities, which indicates one's inability to construct a local-hidden-variable
(LHV) model of the correlations  \cite{Bell64, BrunCPS14}. Such correlations are known  as Bell nonlocal.

Recently, Wiseman \emph{et al}. \cite{WiseJD07, JoneWD07} formalized the concept of Einstein-Podolsky-Rosen (EPR) steering  \cite{EinsPR35, Schr35} and showed that it is  an intermediate type of  nonlocality that sits between entanglement and Bell nonlocality.
EPR steering can be detected by violation of steering inequalities \cite{Reid89,CavaJWR09,KogiSCA15,ZhuHC16}, which indicates one's inability to construct  a hybrid  local hidden variable-local
hidden state (LHV-LHS) model~\cite{Wern89,WiseJD07, JoneWD07}. The asymmetry in the model reflects the asymmetry in the roles played by the two parties  \cite{WiseJD07,BowlVQB14,Hand12}.  Correlations that do not admit an LHV-LHS model  are referred to as EPR nonlocal, which are weaker than Bell-nonlocal correlations and  easier to generate \cite{WiseJD07}. They have received increasing attentions recently because of their intriguing connections with Bell-nonlocal correlations \cite{ChenRCY16,StevB14,QuinVB14,QuinVC15} and  potential applications in quantum information processing, such as quantum key distribution and secure teleportation \cite{Reid09,BranCWS12,Reid13,HeRAR15}.

The  simplest Bell scenario consists of two parties, Alice and Bob, and two binary measurements for each party \cite{Bell64,ClauHSH69,SaunPPS12}. The set of correlations is Bell nonlocal if and only if (iff) it violates the  Clauser-Horne-Shimony-Holt (CHSH) inequality \cite{ClauHSH69}, which is a full-correlation inequality. In addition, there is a simple criterion on  which two-qubit states  can generate such correlations  \cite{HoroHH95,Scar13}. The simplest steering scenario also consists of  two binary measurements for each party, assuming that Bob (the steered party) can only perform projective measurements. Recently, Cavalcanti \emph{et
al}. introduced an analog CHSH inequality and showed
that the set of full correlations (excluding marginal statistics)  is EPR nonlocal iff this inequality is violated, assuming  mutually unbiased measurements for Bob \cite{CavaFFW15}.
However, little is known  about which two-qubit states  are EPR nonlocal in the simplest steering scenario.

 Here we   show  that a two-qubit state can generate  EPR-nonlocal full correlations (excluding marginal statistics) in the simplest steering scenario iff it can generate Bell-nonlocal correlations (note that, in the simplest scenario,
 Bell nonlocality depends on  full correlations, but not on   marginal statistics \cite{Scar13}). When full statistics (including  marginal statistics) is taken into account,  surprisingly, the same scenario can demonstrate  the simplest one-way steering and the strongest hierarchy between steering and Bell nonlocality. In particular, we show that certain two-qubit states are steerable one way in the simplest scenario, but are not steerable the other way (and thus necessarily Bell local) even if one can perform the most sophisticated measurements allowed by quantum mechanics. To illustrate  these intriguing phenomena, several concrete examples are discussed in detail, which are amenable to  demonstration in simple experiments.

In the course of  study, we introduce  restricted LHS  models, and thereby derive a semidefinite program (SDP) for determining the steerability of any bipartite state under given measurements.
 Furthermore, we derive  necessary and sufficient analytical criteria  in several cases of strong theoretical and experimental interest, for example, when   Alice and Bob share a two-qubit state and Alice performs two binary measurements.

Our study settles a fundamental question  on steering in the simplest nontrivial scenario, which may serve as a starting point for exploring steering in  more complicated scenarios. Meanwhile, it  provides valuable insight on the relations between entanglement,  steering, and Bell nonlocality.
It  also helps explain
the relation between steering and incompatibility of observables,  which complements several recent studies \cite{QuinVB14, UolaMG14, UolaBGP15,ZhuHC16}.

The rest of the paper is organized as follows.
In \sref{sec:model}, we introduce the concept of restricted LHS models and thereby propose a SDP for determining the steerability of bipartite states under given measurements of the two parties. In \sref{sec:scenario}, we derive  a necessary and sufficient steering criterion in the simplest nontrivial scenario. In \sref{sec:CSB}, we show that a two-qubit state can generate EPR-nonlocal full correlations in this scenario iff it can generate
Bell-nonlocal (full) correlations. The relations between entanglement,  EPR steering, and Bell nonlocality are then discussed briefly.
In \sref{sec:HierarchySimp}, we provide
a concrete example to show that violation of full-correlation inequalities is
sufficient, but not necessary, for demonstrating steering even in the simplest steering scenario, in sharp contrast with the simplest Bell scenario. This example also manifests a strict hierarchy between steering and Bell nonlocality. In
 \sref{sec:One-Way}, we show that the above steering scenario can demonstrate one-way steering and the hierarchy between steering and Bell nonlocality in the simplest and strongest form.
\Sref{sec:summary} summarizes  this paper.

\section{Steering criteria based on restricted LHS models}
\label{sec:model}
\subsection{Restricted LHS models}

Suppose Alice and Bob share a bipartite state $\rho$ and they can perform measurements in the sets  $M_\rmA$ and $M_\rmB$, respectively, which are referred to as \emph{measurement assemblages} henceforth.  Let $p(a,b|A,B)$ be the probability of obtaining the outcomes $a$ and $b$ when Alice and Bob perform  measurements  $A\in M_\rmA$ and $B\in
M_\rmB$.  Then the state $\rho$  is \emph{steerable}  if the set of probability distributions $p(a,b|A,B)$ does not admit an LHV-LHS model \cite{WiseJD07, JoneWD07},
\begin{equation}
\label{eq:LHVLHS}
 p(a,b|A,B)=\sum_\lambda p_\lambda p(a|A,\lambda)p(b|B,\rho_\lambda).
\end{equation}
 Here, $p(a|A,\lambda)$ for each $A$ and $\lambda$  denotes an  arbitrary probability distribution as does $p_\lambda$, while
$p(b|B,\rho_\lambda)=\tr(\rho_\lambda B_{b})$ denotes the quantum probability of outcome $b$ when Bob performs the measurement $B$  on the state $\rho_\lambda$, where $B_b$ is the effect [element in a positive-operator-valued measure (POVM)] corresponding to  outcome $b$.

If Alice  performs the measurement $A\in M_\rmA$
on the bipartite state~$\rho$ and obtains the outcome $a$, then  Bob's subnormalized reduced state is  $\rho_{a|A}=\tr_\mathrm{A}[(A_{a}\otimes
I)\rho]$, which satisfies $\sum_a\rho_{a|A}=\rho_\mathrm{B}:=\tr_\mathrm{A}(\rho)$.
The set of  states $\{\rho_{a|A}\}_a$ for a given measurement
$A$ is an \emph{ensemble}  for $\rho_\mathrm{B}$, and the whole collection
of ensembles $\{\rho_{a|A}\}_{a,A}$ is a \emph{state assemblage} \cite{Puse13}.
The  state assemblage $\{\rho_{a|A}\}_{a,A}$ is  \emph{unsteerable} if there
exists an  LHS model~\cite{WiseJD07, JoneWD07, QuinVB14,
UolaMG14, UolaBGP15, ZhuHC16},
\begin{equation}\label{eq:LHS}
\rho_{a|A}=\sum_\lambda p_\lambda p(a|A,\lambda)\rho_\lambda,
\end{equation}
 where $p(a|A,\lambda)\geq0$, $\sum_a p(a|A,\lambda)=1$, and $p_\lambda\rho_\lambda$
 form an ensemble for $\rho_\mathrm{B}$. The steerability of the assemblage $\{\rho_{a|A}\}_{a,A}$ can be determined by SDP \cite{Puse13,SkrzNC14}.

 To  elucidate the connection between the LHS model in \eref{eq:LHS} and the LHV-LHS
 model in \eref{eq:LHVLHS}, here we  introduce the  concept of restricted LHS models. Let $\mathcal{B}(\mathcal{H})$ be the space of all operators acting on Bob's  Hilbert space $\mathcal{H}$  and  $\mathcal{V}\leq\mathcal{B}(\mathcal{H})$  a subspace.
 The assemblage  $\{\rho_{a|A}\}_{a,A}$ admits a $\mathcal{V}$-restricted LHS model
 if
 \begin{equation}
 \tr(\Pi \rho_{a|A})=\sum_\lambda p_\lambda p(a|A,\lambda)\tr(\Pi\rho_\lambda)\quad
 \forall\; \Pi\in \mathcal{V},
 \end{equation}
 where $p(a|A,\lambda)\geq0$, $\sum_a p(a|A,\lambda)=1$, and $p_\lambda\rho_\lambda$
 form an ensemble for a quantum state. In that case, the state assemblage  $\{\rho_{a|A}\}_{a,A}$ is  called  $\mathcal{V}$-unsteerable.
 The  assemblage is unsteerable iff it is $\mathcal{B}(\mathcal{H})$-unsteerable. Any
 $\mathcal{V}$-unsteerable
 assemblage is also $\mathcal{W}$-unsteerable if $\mathcal{W}\leq \mathcal{V}$.

 Suppose $\mathcal{R}$ is the space spanned by all the effects $B_b$ in
Bob's measurement assemblage.
Then the set of  probability distributions $p(a,b|A,B)$
 admits an LHV-LHS model iff the state assemblage $\{\rho_{a|A}\}_{a,A}$ is $\mathcal{R}$-unsteerable in view of the  relation
 \begin{equation}
p(a,b|A,B)=\tr(\rho_{a|A}B_b).
 \end{equation}
The existence of such a model does not depend on any detail of  Bob's measurement assemblage except for the  span $\mathcal{R}$.

\subsection{Semidefinite programming for determining steerability}
\label{sec:SDP}

The $\mathcal{V}$-steerability of the assemblage $\{\rho_{a|A}\}_{a,A}$ can be resolved by a similar SDP used for determining the steerability of $\{\rho_{a|A}\}_{a,A}$.
To start with, let us first review the SDP for determining the steerability of the assemblage $\{\rho_{a|A}\}_{a,A}$. Recall that $\{\rho_{a|A}\}_{a,A}$ is steerable iff it does not admit an LHS model as
\begin{equation}
\rho_{a|A}=\sum_\lambda p_\lambda p(a|A,\lambda)\rho_\lambda=\sum_\lambda  p(a|A,\lambda)\sigma_\lambda,
\end{equation}
where $\sigma_\lambda=p_\lambda\rho_\lambda$  form an ensemble of $\rho_\rmB$. Here the conditional probability distributions $p(a|A,\lambda)$ can be replaced by  deterministic conditional probability distributions $D(a|A,\lambda)$, which are extremal.
Then the steerability of the assemblage can be determined by the following SDP \cite{Puse13,SkrzNC14},
\begin{align}
&\mbox{find}\quad \{\sigma_\lambda\} \quad \mbox{subjected to}\nonumber\\
&\sum_\lambda  D(a|A,\lambda)\sigma_\lambda=\rho_{a|A}\quad \forall\; A, a,\nonumber\\
&\tr\biggl(\sum_{\lambda}\sigma_\lambda\biggr)=1,\quad \sigma_\lambda\geq0 \quad \forall\; \lambda.
\end{align}

The assemblage $\{\rho_{a|A}\}_{a,A}$ is $\mathcal{R}$-steerable iff  it cannot be written as follows,
 \begin{equation}
 \tr(\Pi \rho_{a|A})=\sum_\lambda  D(a|A,\lambda)\tr(\Pi\sigma_\lambda)\quad
 \forall \; A, a;\; \forall\; \Pi\in \mathcal{R}.
 \end{equation}
Let $\{\Pi_j\}$ be a basis of $\mathcal{R}$. Then the $\mathcal{R}$-steerability can  be determined by the following SDP,
 \begin{align}
 &\mbox{find}\quad \{\sigma_\lambda\}\quad \mbox{subjected to}\nonumber\\
 &\sum_\lambda  D(a|A,\lambda)\tr(\Pi_j\sigma_\lambda)=\tr(\Pi_j \rho_{a|A})\quad \forall\; A, a, j,\nonumber\\
 &\tr\biggl(\sum_{\lambda}\sigma_\lambda\biggr)=1,\quad \sigma_\lambda\geq0 \quad \forall\; \lambda.
 \end{align}
In this way, the steerability of any bipartite state under given measurements can be determined by SDP.

\section{EPR-nonlocal correlations in the simplest scenario}
\label{sec:scenario}
\subsection{\label{sec:SimpleScenario}The simplest steering scenario}

What is the simplest steering scenario? This question has attracted a lot of attention recently \cite{SaunPPS12, CavaFFW15, ZukoDY15, RoyBMB15, QuanZLF16}. To demonstrate steering, Alice and Bob need to share an entangled state, and the simplest candidate is a two-qubit state. Alice also needs a choice over her measurements, and the simplest  measurement assemblage  consists of  two projective measurements. It is not necessary for Bob to have a choice over different measurements, but the  effects in his measurement assemblage cannot commute with each other pairwise, so the span of these effects has dimension at least 3; the simplest measurement assemblage satisfying this property consists of either two projective measurements or one trine measurement \cite{SaunPPS12}. To summarize, in the simplest steering scenario, Alice and Bob share a two-qubit state, and there are two choices for the measurement setting:
\begin{enumerate}
\item \label{enu:TwoProj} Two  projective measurements for Alice and Bob, respectively.
\item \label{enu:trine} Two  projective measurements for Alice and one trine measurement for Bob.
\end{enumerate}
Here, scenario~2  is singled out by the least complexity cost of 12, where
the complexity cost is  the number of possible patterns of joint detection outcomes that can occur~\cite{SaunPPS12}; see \aref{asec:Hierarchy}. In practice, however, it is usually easier to perform two projective measurements than one trine measurement,  so we shall focus on scenario~1 in the following discussions. Note that the span of effects of any trine measurement
on a qubit can be realized by
two projective measurements, and vice versa. In view of the restricted LHS models discussed above, the two scenarios are actually equivalent as far as determining steerability is concerned, although one scenario may be easier than the other for experimental implementation. Therefore,  many of our conclusions on scenario~1 are also applicable to scenario~2. In addition, similar analysis is applicable when the  projective  measurements of Alice and Bob are replaced by general binary measurements.

\subsection{Necessary and sufficient steering criterion}

The steerability of any two-qubit state under arbitrary given measurements can be determined using the SDP presented in \sref{sec:SDP}. In the simplest steering scenario, we can even derive an analytical criterion.

Consider the two-qubit state
\begin{equation}\label{eq:twoqubit}
\rho=\frac{1}{4}\biggl(I\otimes I+\vec{\alpha}\cdot\vec{\sigma}\otimes I+I\otimes\vec{\beta}\cdot\vec{\sigma}+\sum_{i,j=1}^3t_{ij}\sigma_i\otimes\sigma_j\biggr),
\end{equation}
where $\sigma_j$ for $j=1,2,3$ are three Pauli matrices, $\vec{\sigma}$ is the vector composed of them,  $\vec{\alpha}$
and $\vec{\beta}$ are the Bloch vectors
for Alice and Bob, respectively, and  $T=(t_{ij})$ is the correlation matrix.
Both Alice and Bob can choose two   projective measurements as described by $\{A_1, A_2\}=\{\vec{a}_1\cdot\vec{\sigma},
\vec{a}_2\cdot\vec{\sigma}\}$ and $\{B_1, B_2\}=\{\vec{b}_1\cdot\vec{\sigma},
\vec{b}_2\cdot\vec{\sigma}\}$,
respectively, where $\vec{a}_1, \vec{a}_2, \vec{b}_1, \vec{b}_2$ are unit vectors in dimension~3, referred to as measurement vectors henceforth. The two outcomes of each measurement are denoted by $\pm$.

If Alice obtains outcomes $\pm$ given measurements $m=1,2$,
then the unnormalized reduced states of Bob are given by
\begin{align}
\rho_{\pm|m}&=\frac{1}{2}\tr_\rmA\bigl\{[(I\pm\vec{a}_m\cdot \vec{\sigma})\otimes I]\rho\bigr\}\nonumber \\
&=\frac{1}{4}\bigl[(1\pm\vec{\alpha}\cdot \vec{a}_m)I+\vec{\beta}\cdot\vec{\sigma}\pm\vec{\gamma}_m\cdot\vec{\sigma}\bigr],
\end{align}
where $\gamma_{mj}=\sum_{i=1}^3a_{mi}t_{ij}$.
The span  of effects of Bob's measurements is
$\mathcal{R}=\spa\{I,\vec{b}_1\cdot\vec{\sigma},
\vec{b}_2\cdot\vec{\sigma} \}$, which is three dimensional except in the trivial case in which $\vec{b}_1$ and $\vec{b}_2$ are parallel or antiparallel.
The state $\rho$ is steerable under given measurements iff the assemblage  $\{\rho_{\pm|m}\}_{m=1,2}$ is $\mathcal{R}$-steerable.

Let $\tilde{\rho}_\rmB$ be the orthogonal projection of $\rho_\rmB$  onto $\mathcal{R}$.
Then, $\tilde{\rho}_\rmB$ is  a density matrix whose Bloch vector is the projection of the Bloch vector $\vec{\beta}$ of $\rho_\rmB$ onto the plane spanned by $\vec{b}_1, \vec{b}_2$. Let $\tilde{\rho}_{\pm|m}$ be the orthogonal projection of $\rho_{\pm|m}$  onto $\mathcal{R}$; then $\{\tilde{\rho}_{\pm|m}\}_{m=1,2}$ is an assemblage for $\tilde{\rho}_\rmB$.
To be concrete, we have
\begin{equation}\label{eq:proj}
\tilde{\rho}_\rmB=\frac{1}{2}(I+\tilde{\vec{\beta}}\cdot\vec{\sigma}),\quad \tilde{\rho}_{\pm|m}
=\frac{1}{4}\bigl[(1\pm\vec{\alpha}\cdot \vec{a}_m)I+\tilde{\vec{\beta}}\cdot\vec{\sigma}\pm\tilde{\vec{\gamma}}_m\cdot\vec{\sigma}\bigr],
\end{equation}
where
\begin{equation}\label{eq:proj2}
\tilde{\vec{\beta}}=\sum_{m=1}^2(\vec{\beta}\cdot \vec{b}_m)\vec{b}_m', \quad \tilde{\vec{\gamma}}_m=\sum_{n=1}^2(\vec{\gamma}_m\cdot \vec{b}_n)\vec{b}_n',
\end{equation}
and $\vec{b}_1', \vec{b}_2'$ form the dual basis of  $\vec{b}_1, \vec{b}_2$ in their span, as characterized by the equation $\vec{b}_m'\cdot \vec{b}_n=\delta_{mn}$, assuming $\vec{b}_1, \vec{b}_2$ are  linearly independent.
Note that $\vec{b}_m'= \vec{b}_m$ when $\vec{b}_1$ and $\vec{b}_2$ are orthogonal, that is,  when Bob's measurements are mutually unbiased.
To appreciate the significance of the $\mathcal{R}$-restricted assemblage $\{\tilde{\rho}_{\pm|m}\}_{m=1,2}$, note that it is steerable iff  $\{\rho_{\pm|m}\}_{m=1,2}$ is $\mathcal{R}$-steerable, that is, iff $\rho$ is steerable under given measurements.

If $\tilde{\rho}_\rmB$ is not invertible, then it is   pure,  so the assemblage $\{\tilde{\rho}_{\pm|m}\}_{m=1,2}$ cannot be steerable. Otherwise,
the steerability of $\{\tilde{\rho}_{\pm|m}\}_{m=1,2}$ is equivalent to the incompatibility of the set of steering-equivalent observables  $\{O_{\pm|m}\}_{m=1,2}$ \cite{UolaBGP15} with
\begin{equation}
O_{\pm|m}=\tilde{\rho}_\rmB^{-1/2}\tilde{\rho}_{\pm|m}\tilde{\rho}_\rmB^{-1/2}.
\end{equation}
The latter statement is equivalent to the noncoexistence of the two qubit effects $O_{+|1}$ and $O_{+|2}$, which can be determined analytically \cite{StanRH08,BuscS09, YuLLO10, UolaBGP15}. To be  specific, express $O_{\pm|m}$ as follows:
\begin{align}\label{eq:SEOpar}
O_{\pm|m}=\frac{1}{2}[(1\pm\eta_m)I\pm\vec{r}_m\cdot\vec{\sigma}],
\end{align}
where $\eta_m$ is a real constant, and $\vec{r}_m$ is a real vector in dimension 3, which satisfies $|\eta_m|+|\vec{r}_m|\leq 1$.
Then, $O_{+|1}$ and $O_{+|2}$ are coexistent iff \cite{YuLLO10}
\begin{align}\label{eq:vio}
(1-F_1^2-F_2^2)\Bigl(1-\frac{\eta_1^2}{F_1^2}-\frac{\eta_2^2}{F_2^2}\Bigr)\leq(\vec{r}_1\cdot\vec{r}_2-\eta_1\eta_2)^2,
\end{align}
where
\begin{equation}
F_m=\frac{1}{2}\Bigl(\sqrt{(1+\eta_m)^2-r_m^2}+\sqrt{(1-\eta_m)^2-r_m^2}\Bigr).
\end{equation}
In this way, we can determine analytically  the steerability of any two-qubit state under  two given projective measurements on each side.

We emphasize that all the parameters entering the inequality in \eref{eq:vio} are determined by measurement vectors of Bob and  measurement statistics. So it is straightforward to test this inequality in experiments. To see this, note that
\begin{equation}\label{eq:MeasStatistics}
\begin{aligned}
\langle A_m\rangle&=\vec{\alpha}\cdot \vec{a}_m,\\ \langle B_n\rangle&=\vec{\beta}\cdot \vec{b}_n, \\
\langle A_m B_n\rangle&=\vec{a}_m^\rmT T\vec{b}_n=\vec{\gamma}_m\cdot\vec{b}_n.
\end{aligned}
\end{equation}
Therefore, $\vec{\alpha}\cdot \vec{a}_m$ is determined by measurement statistics;  $\tilde{\vec{\beta}}$ and $\tilde{\vec{\gamma}}_m$ are determined by measurement statistics and Bob's measurement vectors according to \esref{eq:proj2} and \eqref{eq:MeasStatistics} (note that $\vec{b}_1',\vec{b}_2'$ are determined by $\vec{b}_1,\vec{b}_2$). Consequently, $\tilde{\rho}_\rmB$ and $\tilde{\rho}_{\pm|m}$ are completely  determined by measurement vectors of Bob and  measurement statistics, from which our claim follows. In practice, we may also use other tomographic methods for reconstructing $\tilde{\rho}_\rmB$ and $\tilde{\rho}_{\pm|m}$, which may offer some advantage. For example, the maximum-likelihood method \cite{Hrad97,TeoZER11} can ensure  the positivity of the  operators reconstructed. This topic has received little attention in the study of steering, but deserves further study in its own right.

The above analysis is still applicable even if
Alice performs two general binary measurements and Bob performs arbitrary measurements. In addition, the dimension of Alice's Hilbert space can be  larger than 2. To see this, note that the span $\mathcal{R}$ of effects in  Bob's measurement assemblage has dimension 1 to 4. In addition,  Bob's state assemblage $\{\rho_{\pm|m}\}_{m=1,2}$  has two ensembles, each of which consists of two subnormalized qubit states as in the above discussion, and the same holds for the  $\mathcal{R}$-restricted assemblage $\{\tilde{\rho}_{\pm|m}\}_{m=1,2}$. The bipartite state shared by Alice and Bob is steerable by the given measurements iff  $\{\tilde{\rho}_{\pm|m}\}_{m=1,2}$ is steerable.  When  $\dim(\mathcal{R})=4$, we have $\tilde{\rho}_{\pm|m}=\rho_{\pm|m}$; when  $\dim(\mathcal{R})\leq 3$, $\tilde{\rho}_{\pm|m}$ can be computed in a similar way as in the above discussion.
In addition, the  state  cannot be steerable when $\dim(\mathcal{R})\leq 2$, in which case all  effects in  Bob's measurement assemblage commute with each other and thus can be diagonalized simultaneously.

\section{\label{sec:CSB}Coincidence of EPR-nonlocal full correlations and Bell-nonlocal correlations in the simplest scenario}

Although the steering scenario discussed in \sref{sec:scenario} looks  simple, it is  nontrivial to determine the steerability of a generic two-qubit state under optimal measurements. It is  instructive to note that in the simplest Bell scenario,  the CHSH inequality, a full-correlation inequality, is  the only nontrivial Bell inequality  \cite{ClauHSH69, Scar13}.  To address a similar problem in the steering scenario, an analog CHSH inequality was recently proposed by  Cavalcanti \emph{et al}.  in the case
Bob performs two mutually unbiased projective measurements \cite{CavaFFW15}, which was extended to a more general setting later \cite{GirdC16}.  In this section, we propose a simpler analog CHSH inequality whose violation is both necessary and sufficient for demonstrating  EPR-nonlocal full correlations (excluding marginal statistics). Furthermore, we determine the maximum violation of the analog CHSH inequality for any two-qubit state and thereby show that a two-qubit state can generate EPR-nonlocal full correlations in the simplest steering scenario iff it can generate Bell-nonlocal correlations.

\subsection{Analog CHSH inequality for EPR-nonlocal full correlations}

The full correlations between
$A_m$ and $B_n$ for $m,n=1,2$ are given by
the third line in \eref{eq:MeasStatistics}.
The set of full  correlations is Bell nonlocal if it does not admit an LHV model
\cite{Bell64,BrunCPS14,Scar13},
\begin{equation}
\label{eq:LHVsimp}
\langle A_m B_n\rangle=\sum_\lambda p_\lambda \rmE(A_m,\lambda)\rmE(B_n,\lambda), \;\; m,n=1,2.
\end{equation}
Here,
\begin{equation}
\begin{aligned}
\rmE(A_m,\lambda)&=p(+|A_m,\lambda)-p(-|A_m,\lambda),\\ \rmE(B_n,\lambda)&=p(+|B_n,\lambda)-p(-|B_n,\lambda),
\end{aligned}
\end{equation}
where  $p_\lambda$ is an arbitrary probability distribution, while $p(\pm|A_m,\lambda)$ and $p(\pm|B_n,\lambda)$ are arbitrary conditional
 probability distributions.
It turns out that such a model cannot exist iff  these correlations violate
the celebrated  CHSH inequality \cite{ClauHSH69,Scar13},
\begin{align}
\label{eq:CHSH}
\langle A_1B_1\rangle+\langle A_2B_1\rangle+\langle A_1B_2\rangle-\langle A_2B_2\rangle \leq2,
\end{align}
up to relabeling of the measurements and outcomes.

The set of full  correlations is EPR nonlocal if it does not admit an LHV-LHS model \cite{CavaFFW15},
\begin{equation}
\label{eq:LHVLHSc}
\langle A_m B_n\rangle=\sum_\lambda p_\lambda \rmE(A_m,\lambda)\rmE(B_n,\rho_\lambda), \;\; m,n=1,2,
\end{equation}
where
\begin{equation}
\rmE(B_n,\rho_\lambda)=p(+|B_n,\rho_\lambda)-p(-|B_n,\rho_\lambda).
\end{equation}
Compared with the LHV model in \eref{eq:LHVsimp}, here the only difference is that
$p(\pm|B_n,\rho_\lambda)$ are quantum probabilities determined by the Born rule.

 Recently, an analog CHSH inequality was derived in \rcite{GirdC16} based on the earlier work in \rcite{CavaFFW15}, which yields a necessary and sufficient criterion on demonstrating EPR-nonlocal full correlations in the simplest steering scenario.
Here we propose a simpler criterion.
\begin{theorem}\label{thm:GACHSH}
In the simplest steering scenario, the set of full correlations is EPR nonlocal iff	the analog CHSH inequality
\begin{align}\label{eq:GACHSH}
&|\langle (A_1+A_2) B_1\rangle \vec{b}_1'+\langle (A_1+A_2) B_2\rangle \vec{b}_2'|\nonumber\\
&+|\langle (A_1-A_2) B_1\rangle \vec{b}_1'+\langle (A_1-A_2) B_2\rangle \vec{b}_2'|\leq2
\end{align}
is violated,
where $\vec{b}_1', \vec{b}_2'$ form the dual basis of the basis composed of Bob's measurement vectors $\vec{b}_1, \vec{b}_2$ within their span, assuming $\vec{b}_1, \vec{b}_2$ are linearly independent.
\end{theorem}
Unlike the CHSH inequality, the analog CHSH inequality in \eref{eq:GACHSH} is not linear in the correlation functions. Accordingly, the set of EPR local correlations (those correlations that admit LHV-LHS models) does not form a polytope. In addition,  EPR-nonlocal full correlations do not necessarily violate  the CHSH inequality and can be Bell local (that is,  admit LHV models) in certain scenarios. This conclusion is in contrast with the fact that EPR-local correlations can attain the same upper bound~2 of the CHSH inequality as Bell-local correlations \cite{ZukoDY15}. The criterion in \thref{thm:GACHSH} bears a strong resemblance to the criterion on the noncoexistence of two unbiased qubit effects \cite{Busc86,Zhu15IC}; this is not a coincidence, as reflected in the following proof.
\begin{proof}
According to the discussion in \sref{sec:scenario}, the two-qubit state $\rho$ in \eref{eq:twoqubit} is steerable under the measurements $A_1, A_2, B_1, B_2$ iff
the assemblage $\{\tilde{\rho}_{\pm|m}\}_{m=1,2}$ in \eref{eq:proj} is steerable. If $\rho$ is a Bell-diagonal state, then  $\tilde{\rho}_{\pm|m}=(1\pm\tilde{\vec{\gamma}}_m\cdot\vec{\sigma})/4$ given that $\vec{\alpha}=\vec{\beta}=0$. According to \lref{slem:GBCriterion} below, $\{\tilde{\rho}_{\pm|m}\}_{m=1,2}$ is unsteerable iff
\begin{equation}\label{eq:GBCriterionApp}
|\tilde{\vec{\gamma}}_1+\tilde{\vec{\gamma}}_2|+|\tilde{\vec{\gamma}}_1-\tilde{\vec{\gamma}}_2|\leq 2.
\end{equation}
Alternatively, this equation also follows from \eref{eq:vio}.
According to \esref{eq:proj2} and \eqref{eq:MeasStatistics}, \eref{eq:GBCriterionApp} is equivalent to the
analog CHSH inequality in \eref{eq:GACHSH}.
 For Bell-diagonal states, the marginal statistics are completely random, so the existence of an LHV-LHS model for full correlations is equivalent to  the existence of an LHV-LHS model for  full statistics. Therefore, the set of full correlations is EPR local iff the analog CHSH inequality is satisfied.

When $\rho$ is not a Bell-diagonal state, let $\varrho$ be the Bell  diagonal  state with the same correlation matrix as $\rho$. Then, $\varrho$  and $\rho$ generate the same full correlations. In particular, the set of full correlations
generated by $\rho$ is EPR local iff that generated by $\varrho$ is EPR local. Now Theorem~1 follows from the above conclusion on Bell-diagonal states.
\end{proof}

In general, violation of
\eref{eq:GBCriterionApp} is sufficient, but not necessary, for witnessing steering of the assemblage  $\{\tilde{\rho}_{\pm|m}\}_{m=1,2}$ (cf.~\ssref{sec:HierarchySimp} and \ref{sec:One-Way}). Accordingly, violation of the analog CHSH inequality is sufficient, but not necessary, for demonstrating steering even in the simplest steering scenario.

\begin{lemma}\label{slem:GBCriterion}
Let $\{\rho_{\pm|m}\}_{m=1,2}$ be a qubit assemblage with
\begin{equation}
\rho_{\pm|m}=\frac{1}{4}[(1\pm\nu_m)I+\vec{\beta}\cdot\vec{\sigma}\pm\vec{\gamma}_m\cdot\vec{\sigma}].
\end{equation}
If $\{\rho_{\pm|m}\}_{m=1,2}$ is unsteerable, then
\begin{equation}
\label{eq:GBCriterion}
|\vec{\gamma}_1+\vec{\gamma}_2|+|\vec{\gamma}_1-\vec{\gamma}_2|\leq 2.
\end{equation}
If $\nu_1=\nu_2=0$ and $\vec{\beta}=0$, then $\{\rho_{\pm|m}\}_{m=1,2}$ is unsteerable iff \eref{eq:GBCriterion} is satisfied.
\end{lemma}
\begin{remark}
	When $\vec{\beta}=0$, the lemma follows from the result on the coexistence of two qubit effects \cite{Busc86,Zhu15IC}.
\end{remark}

\begin{proof}
If $\{\rho_{\pm|m}\}_{m=1,2}$ is unsteerable, then any assemblage obtained from $\{\rho_{\pm|m}\}_{m=1,2}$ by applying a unitary transformation is also unsteerable. In addition, permuting the elements in each ensemble in the assemblage does not change steerability.  Therefore, the assemblage $\{[(1\pm\nu_m)I-(\vec{\beta}\cdot\vec{\sigma}\pm\vec{\gamma}_m\cdot\vec{\sigma})]/4\}_{m=1,2}$ is unsteerable, and so is the assemblage $\{\rho'_{\pm|m}\}_{m=1,2}$ with $\rho'_{\pm|m}=[(1\mp\nu_m)I-\vec{\beta}\cdot\vec{\sigma}\pm\vec{\gamma}_m\cdot\vec{\sigma}]/4$. Consequently, the assemblage $\{(\rho_{\pm|m}+\rho'_{\pm|m})/2\}_{m=1,2}=\{(I\pm\vec{\gamma}_m\cdot\vec{\sigma})/4\}_{m=1,2}$ is unsteerable. In view of the relation between steerability and joint measurability \cite{UolaBGP15,ZhuHC16}, this means that   the measurement assemblage $\{(I\pm\vec{\gamma}_m\cdot\vec{\sigma})/2\}_{m=1,2}$ is compatible, which is the case  iff \eref{eq:GBCriterion} is satisfied according to \rcite{Busc86}.

If $\nu_1=\nu_2=0$ and $\vec{\beta}=0$, then $\{\rho_{\pm|m}\}_{m=1,2}$ is unsteerable iff  the measurement assemblage $\{(I\pm\vec{\gamma}_m\cdot\vec{\sigma})/2\}_{m=1,2}$ is compatible, which is the case iff \eref{eq:GBCriterion} is satisfied~\cite{Busc86}.
\end{proof}

If Bob's measurements are mutually unbiased, that is, $\vec{b}_1\bot\vec{b}_2$, then $\vec{b}_1'=\vec{b}_1, \vec{b}_2'=\vec{b}_2$, so that \eref{eq:GACHSH} reduces to the special analog CHSH inequality
\begin{align}\label{eq:ACHSH}
&\sqrt{\langle(A_1+A_2)B_1\rangle^2+\langle(A_1+A_2)B_2\rangle^2}\nonumber\\
&+\sqrt{\langle(A_1-A_2)B_1\rangle^2+\langle(A_1-A_2)B_2\rangle^2}\leq2,
\end{align}
which was first derived by Cavalcanti \emph{et al}. \cite{CavaFFW15} (and called the analog CHSH inequality in their paper).
Comparison of \esref{eq:CHSH} and \eqref{eq:ACHSH} may give the impression that it is easier to generate EPR-nonlocal full correlations than Bell-nonlocal (full) correlations.

\subsection{A surprising coincidence}

\begin{theorem}\label{thm:BellSteer}
A two-qubit state  can generate Bell-nonlocal
 correlations  in the simplest nontrivial scenario  iff it can generate EPR-nonlocal full correlations.
\end{theorem}
This theorem follows from \thref{thm:ACHSH} below. An alternative proof is presented in \aref{asec:CSBProofAlt}.  Shortly after our original draft was posted (arXiv:1601.00962v1), Girdhar and  Cavalcanti  derived a similar result independently (arXiv:1601.01703), which has been published by now \cite{GirdC16}.

\begin{theorem}\label{thm:ACHSH}
The maximal violation $S$ of the analog CHSH inequality by any two-qubit state with correlation matrix $T$  is equal to the maximal violation of the CHSH inequality, namely, $S=2\sqrt{\lambda_1+\lambda_2}$, where
$\lambda_1, \lambda_2$ are the two largest eigenvalues of $TT^\rmT$.
Both inequalities are violated iff  $\lambda_1+\lambda_2>1$.
\end{theorem}
Here the maximal violation $S$ is defined as the maximum of the left hand side of \eref{eq:GACHSH} over $A_1, A_2, B_1, B_2$. Alternatively, $S$ may be defined as the maximum of the left hand side of \eref{eq:ACHSH}, since the maximum in the former case can be attained by mutually unbiased measurements for Bob, as shown in the following proof. Strictly speaking, $S$  represents the maximal violation only when it is larger than 2. To avoid verbosity, we shall not mention this subtlety again in the rest of this paper.

\begin{proof}
 According to  Horodecki \emph{et al.} \cite{HoroHH95} (see also Ref.~\cite{Scar13}), the violation of the CHSH inequality   satisfies\begin{align}\label{eq:CHSHprf}
&\langle A_1B_1\rangle+\langle A_2B_1\rangle+\langle A_1B_2\rangle-\langle A_2B_2\rangle\nonumber\\
&=(\vec{a}_1^\rmT+\vec{a}_2^\rmT)T\vec{b}_1+(\vec{a}_1^\rmT-\vec{a}_2^\rmT)T\vec{b}_2\nonumber\\
&\leq|T^\rmT(\vec{a}_1+\vec{a}_2)|+|T^\rmT(\vec{a}_1-\vec{a}_2)|\leq2\sqrt{\lambda_1+\lambda_2},
\end{align}
where the first inequality is saturated when $\vec{b}_1$ and $\vec{b}_2$ align with $T^\rmT(\vec{a}_1+\vec{a}_2)$ and $T^\rmT(\vec{a}_1-\vec{a}_2)$, respectively, and the second one is saturated when $\vec{a}_1$ and $\vec{a}_2$ are
 eigenvectors  of $TT^\rmT$ with the two largest eigenvalues \cite{QuanZLF16}.

The violation of the analog CHSH inequality reads
\begin{align}\label{eq:ACHSHprf}
&\quad \bigl|\langle (A_1+A_2) B_1\rangle \vec{b}_1'+\langle (A_1+A_2) B_2\rangle \vec{b}_2'\bigr|\nonumber\\
&\quad +\bigl|\langle (A_1-A_2) B_1\rangle \vec{b}_1'+\langle (A_1-A_2) B_2\rangle \vec{b}_2'\bigr|\nonumber\\
&=\bigl|[(\vec{a}_1^\rmT+\vec{a}_2^\rmT)T \vec{b}_1] \vec{b}_1'+[(\vec{a}_1^\rmT+\vec{a}_2^\rmT)T\vec{b}_2] \vec{b}_2'\bigr|\nonumber\\
&\quad+\bigl|[(\vec{a}_1^\rmT-\vec{a}_2^\rmT)T \vec{b}_1] \vec{b}_1'+[(\vec{a}_1^\rmT-\vec{a}_2^\rmT)T\vec{b}_2] \vec{b}_2'\bigr|\nonumber\\
&\leq|T^\rmT(\vec{a}_1+\vec{a}_2)|+|T^\rmT(\vec{a}_1-\vec{a}_2)|\leq 2\sqrt{\lambda_1+\lambda_2}.
\end{align}
Here the first inequality is saturated when $\vec{b}_1$ and $\vec{b}_2$ form a basis in the span of $T^\rmT\vec{a}_1$ and $T^\rmT\vec{a}_2$; it suffices to consider mutually unbiased measurements for Bob, in agreement with the discussion on restricted LHS models.  The second one is saturated under the same condition as that in  \eref{eq:CHSHprf}.
Therefore, the CHSH and analog CHSH inequalities can be violated iff $\lambda_1+\lambda_2>1$.
\end{proof}

What we  have demonstrated in the above proof is actually stronger than stated
  in \thref{thm:ACHSH}: the maximal violations of the CHSH and analog CHSH inequalities
 for fixed measurements of Alice are also equal.
 However,  there is a crucial difference in Bob's measurements required to saturate the CHSH and analog CHSH inequalities, as reflected in  \thref{thm:ACHSHM} below. It should also be pointed out that \thref{thm:ACHSH} does not imply the equivalence of Bell-local (full) correlations and EPR-local full correlations under given measurements in the simplest scenario.

 \begin{theorem}\label{thm:ACHSHM}
 	Suppose both Alice and Bob
 	can only perform mutually unbiased measurements.
 	The maximal violation of the analog CHSH inequality by any two-qubit state with correlation matrix $T$ is still
 	equal to $S=2\sqrt{\lambda_1+\lambda_2}$, where
 	$\lambda_1, \lambda_2$ are the two largest eigenvalues of $TT^\rmT$.
 	The maximal violation of the CHSH inequality is equal to $S_\rmM=\sqrt{2}(\sqrt{\lambda_1}+\sqrt{\lambda_2})$.
 \end{theorem}
\begin{remark}
In the steering scenario, the measurements of Alice (the steering party) are usually not characterized; in the Bell scenario, the measurements of both Alice and Bob are not characterized. However, to demonstrate nonlocality in experiments, precise characterization and implementation of these measurements are essential. Therefore, \thref{thm:ACHSHM} is instructive to experimental demonstration of the two forms of nonlocality.
\end{remark}

\begin{proof}
	The conclusion on the analog CHSH inequality is clear from the proof of Theorem~3. Concerning the CHSH inequality, since $A_1$ and $A_2$ are mutually unbiased and so are  $B_1$ and $B_2$, we have   $\vec{a}_1\bot \vec{a}_2$ and  $\vec{b}_1\bot\vec{b}_2$. Consequently, \eref{eq:CHSHprf} reduces to
	\begin{align}\label{eq:CHSHMprf}
	&\langle A_1B_1\rangle+\langle A_2B_1\rangle+\langle A_1B_2\rangle-\langle A_2B_2\rangle\nonumber\\ &=\sqrt{2}(\vec{c}_1^\rmT T\vec{b}_1+\vec{c}_2^\rmT T\vec{b}_2)\leq\sqrt{2}(\sqrt{\lambda_1}+\sqrt{\lambda_2}),
	\end{align}
	where $\vec{c}_1=(\vec{a}_1+\vec{a}_2)/\sqrt{2}$ and $\vec{c}_2=(\vec{a}_1-\vec{a}_2)/\sqrt{2}$ are orthonormal. The inequality follows from the variational characterization of singular values \cite{Bhat97book}; it is  saturated when $\vec{c}_1, \vec{c}_2$ ($\vec{b}_1, \vec{b}_2$) are the left (right) singular vectors corresponding to the two largest singular values of $T$.\end{proof}
\begin{remark}
	Note that $\sqrt{2}(\sqrt{\lambda_1}+\sqrt{\lambda_2})\leq 2\sqrt{\lambda_1+\lambda_2}$
	and that the inequality is saturated iff $\lambda_1=\lambda_2$.	Therefore, the maximal violation of the CHSH inequality is reduced under the restriction to mutually unbiased measurements whenever $\lambda_1\neq \lambda_2$.

	\Thsref{thm:ACHSH} and \ref{thm:ACHSHM}
	 reflect  a crucial difference between the optimal measurements in the simplest steering scenario and that in the simplest Bell scenario. In the former case,  it is not necessary
	to align Bob's measurements as long as the span of his measurement vectors
	is the same as that of  $T^\rmT\vec{a}_1$ and $T^\rmT\vec{a}_2$; it suffices to consider mutually unbiased measurements for both Alice and Bob to attain the maximal violation of the analog CHSH inequality.
	In the latter case, by contrast,
	the  optimal measurements are much more restricted, and it is usually impossible to attain the maximal violation of the CHSH inequality if their measurements  are both mutually unbiased.
\end{remark}

\subsection{Relations between entanglement, steering and Bell nonlocality}
\label{Srelation}
\begin{figure}[tb]
 	\centering
 	\includegraphics[scale=0.28]{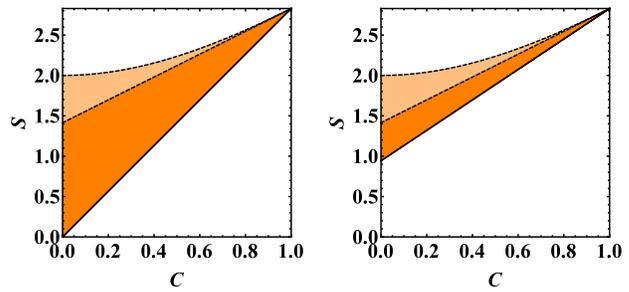}
 	\caption{\label{fig:SC}(color online).
 		Ranges of values of the steering measure $S$ (the common maximal violation of  the CHSH and
 		analog CHSH inequalities) for given concurrence.  Left: General two-qubit states  by \eref{eq:SC}; Right: Entangled
 		Bell-diagonal states by \eref{eq:SCBD}. For comparison,
 		the ranges of values of $S_\rmM$ are rendered in  dark orange (dark gray), where $S_\rmM$
 		is  the maximal violation of the CHSH inequality when both parties  can only perform mutually unbiased
 		measurements; cf.~\esref{eq:SCM} and \eqref{eq:SCMBD}.  In each plot, the lower boundaries  (solid lines) of  $S_\rmM$ and $S$ coincide, but the upper boundaries (dashed lines) are different. }
 \end{figure}

The relation between the maximal violation of the CHSH inequality and the concurrence $C$ for two-qubit states \cite{Woot98} was determined in  \rcite{VersW02}. By \thref{thm:ACHSH}, the same result applies to the analog CHSH inequality:
\begin{equation}\label{eq:SC}
2\sqrt{2}C\leq S\leq 2\sqrt{1+C^2}.
\end{equation}
The relation between $S_\rmM$ and $C$ can be derived similarly,
\begin{equation}\label{eq:SCM}
2\sqrt{2}C\leq S_\rmM\leq \sqrt{2}(1+C).
\end{equation}
Both upper bounds are saturated by pure states and rank-2 Bell-diagonal states,  while lower bounds are saturated by convex combinations of the singlet and a suitable product state (cf.~\sref{sec:HierarchySimp}). By contrast,  for entangled Bell-diagonal states, we have
\begin{align}
\frac{2\sqrt{2}}{3}(1+2C)&\leq S\leq 2\sqrt{1+C^2},\label{eq:SCBD}\\
\frac{2\sqrt{2}}{3}(1+2C)&\leq S_\rmM\leq \sqrt{2}(1+C).\label{eq:SCMBD}
\end{align}
Here \eref{eq:SCBD} is derived in
\rcite{QuanZLF16}; \eref{eq:SCMBD} follows from a similar recipe.  Both lower bounds  are saturated by Werner states, while upper bounds  by rank-2 Bell-diagonal states. The relations between $S, S_\rmM$ and $C$ are illustrated in \fref{fig:SC}.

As an implication of \esref{eq:SC} and \eqref{eq:SCM}, any two-qubit  state with $C>1/\sqrt{2}$
can violate the analog CHSH inequality and is thus steerable. To violate the CHSH inequality under mutually unbiased measurements, it must have concurrence at least $C>\sqrt{2}-1$.   So not all entangled pure states can violate the CHSH inequality in this scenario, although they can under optimal measurements.
The restriction to mutually unbiased measurements severely limits the capability of two-qubit states in generating Bell-nonlocal full correlations, in sharp contrast with the steering scenario, in which this is not a limitation. This observation is instructive for clarifying the distinction between the two forms of nonlocality, which is quite relevant to experimental demonstration. It is also of intrinsic interest to understanding the relation between nonlocality and incompatibility of observables.

\section{\label{sec:HierarchySimp}Strict Hierarchy between steering and Bell nonlocality in the simplest scenario}

In this section, we provide a concrete example of two-qubit states that do not violate the analog CHSH inequality but  are nevertheless  steerable in the simplest steering scenario, that is, two projective measurements for both Alice and Bob. Our example shows that violation of full-correlation inequalities is not necessary for demonstrating steering, even in the simplest steering scenario, once full statistics is taken into account. This conclusion is in sharp contrast with the simplest Bell scenario, in which case violation of the CHSH inequality, a full-correlation inequality, is necessary  for demonstrating Bell nonlocality \cite{ClauHSH69,Scar13}.
In this way, our example  demonstrates a strict hierarchy between steering and Bell nonlocality in the simplest scenario.

Consider the two-qubit state $\rho$ in \eref{eq:twoqubit} with
\begin{equation}
\vec{\alpha}=(0,0,1-s), \quad \vec{\beta}=0, \quad T=-\diag(s,s,s),\quad
\end{equation}
where $0\leq s\leq 1$. More explicitly, $\rho$ has the form
\begin{equation}
\label{eq:hier}
\rho=\frac{1}{4}[
I\otimes I+(1-s) \sigma_3\otimes I-s(\sigma_1\otimes \sigma_1+\sigma_2\otimes \sigma_2+\sigma_3\otimes \sigma_3)].
\end{equation}
The state is a convex combination of the singlet and a product state,
\begin{equation}
\rho=s(|\Psi_-\rangle\langle\Psi_-|)+(1-s)(|0\rangle\langle 0|)\otimes \frac{I}{2} ,
\end{equation}
where $|\Psi_-\rangle=(|01\rangle-|10\rangle)/\sqrt{2}$. The  reduced states of Alice and Bob are given by $\rho_\rmA=[I+(1-s) \sigma_3]/2$ and $\rho_\rmB=I/2$.
The eigenvalues of $\rho$ are
\begin{equation}
0, \quad \frac{1-s}{2},\quad  \frac{1+s\pm\sqrt{1-2s+5s^2}}{4}.
\end{equation}
The partial transpose of $\rho$ reads
\begin{equation}
\frac{1}{4}[I\otimes I+(1-s) \sigma_3\otimes I-s(\sigma_1\otimes \sigma_1-\sigma_2\otimes \sigma_2+\sigma_3\otimes \sigma_3)],
\end{equation}
whose eigenvalues are
\begin{equation}
\frac{1}{2}, \quad  \frac{s}{2},\quad  \frac{1-s\pm\sqrt{1-2s+5s^2}}{4}.
\end{equation}
The negativity and concurrence of $\rho$ read
\begin{align}
N(\rho)=\frac{-1+s+\sqrt{1-2s+5s^2}}{2},\quad
C(\rho)=s.
\end{align}
Therefore, $\rho$ is entangled iff  $0< s\leq 1$.

According to \thref{thm:ACHSH}, the maximum violation of the (analog) CHSH inequality by $\rho$ reads
\begin{equation}\label{eq:SnoCHSH}
S(\rho)=2\sqrt{2}s=2\sqrt{2}C(\rho).
\end{equation}
So, $\rho$ can generate EPR-nonlocal full correlations in the simplest steering scenario
iff $\sqrt{2}/2<s\leq 1$. According to \thref{thm:ACHSHM}, the same formula still applies under the restriction to  mutually unbiased measurements, that is,
\begin{equation}\label{SMnoCHSH}
S_\rmM(\rho)=2\sqrt{2}s=2\sqrt{2}C(\rho).
\end{equation}
The coincidence $S_\rmM(\rho)=S(\rho)$ follows from the fact that all singular values of the correlation matrix $T$ are equal. It is worth pointing out that $S(\rho)$ and $S_\rmM(\rho)$ saturate the lower bounds in \esref{eq:SC} and \eqref{eq:SCM}, respectively.

\begin{figure}[tb]
	\centering
	\includegraphics[scale=0.6]{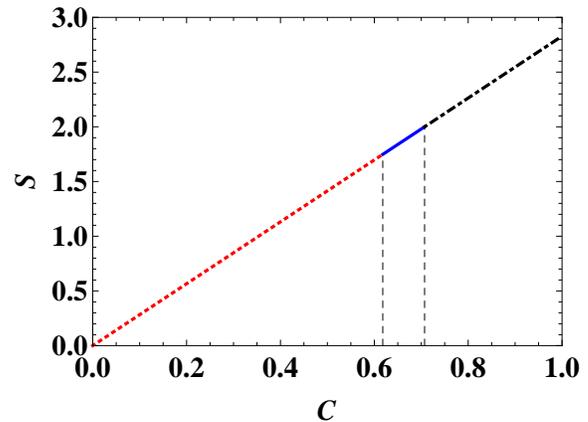}
	\caption{\label{fig:Hierarchy}(color online).
		Hierarchy between entanglement, EPR steering, and Bell nonlocality demonstrated by the family of states in \eref{eq:hier}, which is parametrized by $s$. The oblique line depicts the relation between concurrence $C$ (equal to the parameter $s$) and the maximum violation $S$ of the (analog) CHSH inequality. The black region (dot-dashed line) corresponds to states that can generate Bell-nonlocal correlations or EPR-nonlocal full correlations in the simplest scenario. States corresponding to the blue region sandwiched between the two thin dashed lines are steerable in the same scenario, but cannot generate Bell-nonlocal correlations or EPR-nonlocal full correlations.	
	}
\end{figure}

Suppose Alice can perform two projective measurements  $\vec{a}_m\cdot\vec{\sigma}$ for $m=1,2$. Then the assemblage $\{\rho_{\pm|m}\}$ of Bob induced by Alice has the form
\begin{align}
\rho_{\pm|m}&
=\frac{1}{4}\bigl\{[1\pm(1-s) a_{m3}]I\mp s\vec{a}_m\cdot\vec{\sigma}\bigr\}.
\end{align}
If Bob's measurement vectors are parallel to that of Alice's, that is, $\vec{b}_m=\vec{a}_m$ for $m=1,2$, then
 $\rho_\rmB$ and $\rho_{\pm|m}$ belong to
$\mathcal{R}=\spa\{I,\vec{b}_1\cdot\vec{\sigma},
\vec{b}_2\cdot\vec{\sigma} \}$,  so we have $\tilde{\rho}_\rmB=\rho_\rmB$ and $\tilde{\rho}_{\pm|m}=\rho_{\pm|m}$. The steering-equivalent observables of $\tilde{\rho}_{\pm|m}$ take on the form
\begin{align}
O_{\pm|m}&
=\frac{1}{2}\bigl\{[1\pm(1-s) a_{m3}]I\mp s\vec{a}_m\cdot\vec{\sigma}\bigr\}.
\end{align}
Therefore, $\rho$ is steerable by the given measurements iff the inequality in \eref{eq:vio} is violated, where  $\vec{r}_m= -s \vec{a}_m$ and $\eta_m=(1-s) a_{m3}$ for $m=1,2$.

Choose  $\vec{a}_1=\vec{b}_1=(1,0,0)^\rmT$ and $\vec{a}_2=\vec{b}_2=(0,0,1)^\rmT$; then $\eta_1=0$, $\eta_2=1-s, r_1=r_2=s$, $\vec{r}_1\cdot\vec{r}_2=0$, $F_1=\sqrt{1-s^2}$, and $F_2=\sqrt{1-s}$. According to \eref{eq:vio}, the state $\rho$ is steerable whenever
$s(s+s^2-1)>0$ , that is, $s>(\sqrt{5}-1)/2$. If
\begin{equation}
\frac{1}{2}(\sqrt{5}-1)<s\leq \frac{\sqrt{2}}{2},
\end{equation}
then the state $\rho$ is steerable in the simplest steering scenario, but cannot violate the CHSH or  analog CHSH inequality, as illustrated in  \fref{fig:Hierarchy}. In this way, the state  demonstrates a strict hierarchy between steering and Bell nonlocality in the simplest scenario.

As a side remark, our example  disproves the claim in \rcite{CavaFFW15} that violation of the analog CHSH inequality (when Bob's measurements are mutually unbiased) is both necessary and sufficient for demonstrating steering in the simplest steering scenario. The lapse in \rcite{CavaFFW15} appears in the paragraph above Eq.~(10) there, where they claimed  that the number of free parameters in the set of probability distributions $\mathbf{P}_{\mathcal{AB}}$ is 4 instead of 8; cf.~similar result on the Bell scenario \cite{Scar13}. Note that the constraints on the probability distributions  considered  in \rcite{CavaFFW15} are not independent. By contrast, the number of free parameters characterizing full correlations is only 4. Therefore,  an LHV-LHS model for  full correlations is not \emph{a priori} sufficient for guaranteeing an LHV-LHS model for  full statistics.

\section{Simplest and strongest one-way steering}
\label{sec:One-Way}
In this section, we reveal one-way steering in the  simplest and strongest form.  Building on the previous work in \rcite{BowlHQB16}, we show that certain two-qubit states are steerable one way in the simplest steering scenario, but are not steerable the other way even under the most sophisticated measurements allowed by quantum mechanics.
The following example also demonstrates the strongest hierarchy between steering and Bell nonlocality.
%If marginal statistics is ignored in the simplest steering,

Consider the  two-qubit state \cite{BowlHQB16},
\begin{align}\label{eq:OneWayStates}
&\rho(p,\theta)=p (|\psi(\theta)\rangle \langle \psi(\theta)|)+(1-p)\Bigl[ \frac{I}{2} \otimes\rho_\rmB(\theta)\Bigr]\nonumber \\
&=\frac{1}{4}\Bigl\{I\otimes I +p\cos(2\theta) \sigma_3\otimes I+\cos(2\theta) I\otimes \sigma_3\nonumber\\
&\quad +p\bigl[\sin(2\theta)\sigma_1\otimes \sigma_1-\sin(2\theta)\sigma_2\otimes \sigma_2+\sigma_3\otimes \sigma_3\bigr]\Bigr\},
\end{align}
where $|\psi(\theta)\rangle=\cos\theta|00\rangle +\sin\theta|11\rangle$ with $\sin(2\theta)\neq 0$, and
\begin{align}
 \rho_\rmB(\theta)&=\tr_\rmA(|\psi(\theta)\rangle \langle \psi(\theta)|)=\frac{1}{2}[I+\cos(2\theta)\sigma_3]\nonumber\\
 &=\cos^2\theta (|0\rangle\langle 0|)+\sin^2\theta (|1\rangle\langle1|).
\end{align}
The  two-qubit state $\rho(p,\theta)$ has correlation matrix
\begin{equation}
T=\diag\bigl(p\sin(2\theta), -p\sin(2\theta),p\bigr),
\end{equation}
so it  can violate the analog CHSH inequality iff
 \begin{equation}\label{eq:ACHSHoneWay}
p^2\bigl[1+\sin^2(2\theta)\bigr]>1.
 \end{equation}

According to \rcite{BowlHQB16},  the state $\rho(p,\theta)$ is unsteerable from Bob to Alice  by arbitrary projective measurements and hence cannot violate any Bell inequality when
\begin{equation}\label{eq:UnsteerCon}
\cos^2(2\theta)\geq \frac{2p-1}{(2-p)p^3}.
\end{equation}
 In addition,   the state is steerable from Alice to Bob by two projective measurements  whenever $p>1/\sqrt{2}$. However, the steerability  was not inferred  directly from steering criteria, instead it is inferred  from violation of the CHSH inequality after applying a suitable local filtering operation.  Also, the result presented in \rcite{BowlHQB16} does not \emph{a priori} imply  that the state with  $p>1/\sqrt{2}$ is steerable in the simplest steering scenario under consideration here. Nevertheless,  we  show that our steering criterion derived in \sref{sec:scenario} can certify the simplest one-way steering, which is easy to verify in experiments.

\begin{figure}[tb]
 	\centering
 	\includegraphics[scale=0.58]{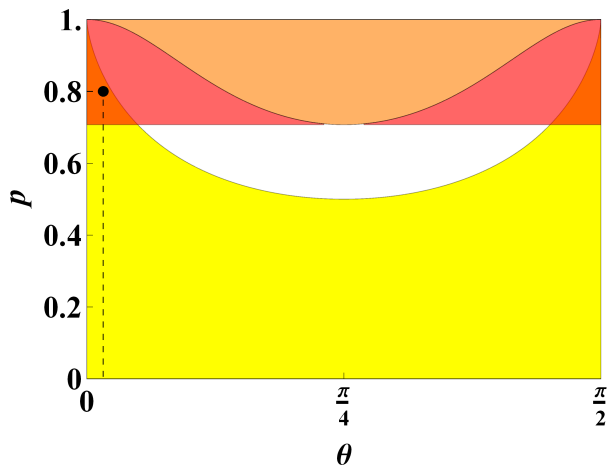}
 	\caption{\label{fig:OW}(color online). States that can demonstrate the simplest and strongest one-way steering. Each parameter point coordinated by $p$ and $\theta$ defines a two-qubit state according to \eref{eq:OneWayStates}. The orange (intermediate gray) region  corresponds to states that can violate the (analog) CHSH   inequality. The red (dark gray) region corresponds to states that are steerable from Alice to Bob in the simplest steering scenario, but that  cannot violate the (analog) CHSH   inequality. States in the yellow (light gray) region are not steerable from Bob to Alice by arbitrary projective measurements. States in the intersection of the red region and the yellow region (including the black dot corresponding to $p=0.8$ and $\theta=0.05$) are the most peculiar because they can  demonstrate the simplest and strongest one-way steering.  }
 \end{figure}

Suppose the projective  measurements of Alice and Bob are determined by the measurement vectors
\begin{equation}\label{eq:OneWayMeasVec}
\vec{a}_1=\vec{b}_1=(1,0,0)^\rmT,\quad \vec{a}_2=\vec{b}_2=(0,0,1)^\rmT.
\end{equation}
Then, the $\mathcal{R}$-restricted state assemblage of Bob is given by
\begin{equation}
\begin{aligned}
\tilde{\rho}_{\pm|1}&=\rho_{\pm|1}
=\frac{1}{4}[I+\cos(2\theta)\sigma_3\pm 2p\sin\theta\cos\theta \sigma_1]\\
&\rep\frac{1}{2}\begin{pmatrix}
\cos^2\theta & \pm p\sin\theta\cos\theta\\ \pm p\sin\theta\cos\theta &\sin^2\theta
\end{pmatrix},\\
\tilde{\rho}_{\pm|2}&=\rho_{\pm|2}=\frac{1}{4}\bigl\{[1\pm p\cos(2\theta)]I+[\cos(2\theta)\pm p]\sigma_3\bigr\}\\
&\rep\frac{1}{2}\diag\bigl((1\pm p)\cos^2\theta, (1\mp p)\sin^2\theta\bigr).
\end{aligned}
\end{equation}
The corresponding steering-equivalent observables  read
\begin{equation}
O_{\pm|1}=\frac{1}{2}(I\pm p\sigma_1),\quad O_{\pm|2}=\frac{1}{2}(I\pm p\sigma_3).
\end{equation}
So,  $\rho$ is steerable from Alice to Bob in the simplest steering scenario whenever $p> 1/\sqrt{2}$ according to the steering criterion in \eref{eq:vio} or \eqref{eq:GBCriterion} (cf.~\rcite{Busc86}). This condition turns out to  be also necessary.  To see this, note that $\rho(p,\theta)$ is obtained from the Bell-diagonal state $\rho(p,\pi/4)$ (equivalent to a Werner state under a local unitary transformation) by applying local filtering operation on Bob's side. In addition $\rho(p,\pi/4)$ is steerable by two projective measurements iff $p> 1/\sqrt{2}$ \cite{QuanZLF16}. Therefore, $\rho(p,\theta)$ is steerable from Alice to Bob in the simplest steering scenario iff
\begin{equation}
p> \frac{1}{\sqrt{2}},
\end{equation}
which is usually much easier to satisfy than the condition in \eref{eq:ACHSHoneWay} required for violating the analog CHSH inequality.
If, in addition, \eref{eq:UnsteerCon} is satisfied ($p=0.8$ and $\theta=0.05$, for example), then  the state $\rho(p,\theta)$ demonstrates the simplest one-way steering and the strongest hierarchy between steering and Bell nonlocality under projective measurements; see \fref{fig:OW}.

Next, we strengthen the above result  by considering POVMs. Consider the state \cite{BowlHQB16}
\begin{align}\label{eq:OWSpovm}
&\frac{1}{2}[\rho(p,\theta)+\rho_\rmA(p,\theta)\otimes (|0\rangle\langle 0|)]\nonumber\\
&=\frac{1}{4}\bigl[I\otimes I +p\cos(2\theta) \sigma_3\otimes I+\cos^2\theta I\otimes \sigma_3\nonumber\\
&\quad +p\cos\theta(\sin\theta\sigma_1\otimes \sigma_1-\sin\theta\sigma_2\otimes \sigma_2+\cos\theta\sigma_3\otimes \sigma_3)\bigr],
\end{align}
where
\begin{equation}
\rho_\rmA(p,\theta)=\tr_\rmB[\rho(p,\theta)]=\frac{1}{2}[I+p\cos(2\theta)\sigma_3].
\end{equation}
It has correlation matrix
\begin{equation} T=\diag\bigl(p\sin\theta\cos\theta,-p\sin\theta\cos\theta,p\cos^2\theta\bigr),
\end{equation}
so  it  can never violate the analog CHSH inequality. In addition,  it is not steerable from Bob to Alice by arbitrary POVMs
when \eref{eq:UnsteerCon} is satisfied,
while it is steerable from Alice to Bob by two projective measurements for a certain parameter range satisfying \eref{eq:UnsteerCon} \cite{BowlHQB16}. Again, this conclusion does not \emph{a priori} imply  that the state is steerable from Alice to Bob in the simplest steering scenario. Fortunately, our steering criteria can certify the simplest one-way steering as in the previous case. For example,  when $p=0.825$ and $\theta=0.020$,  the state in \eref{eq:OWSpovm} is not steerable from Bob to Alice by arbitrary POVMs. However, it  is steerable from Alice to Bob under the same  measurement setting as specified in \eref{eq:OneWayMeasVec}, in which case the  inequality in \eref{eq:vio} is violated by 0.021.

In view of the discussion in  \sref{sec:SimpleScenario}, to demonstrate one-way steering proposed above, Bob may also perform a trine measurement instead of two projective measurements. One particular  trine measurement takes on the form $\{(I+\vec{c}_j\cdot \vec{\sigma})/3\}_{j=1,2,3}$, where
\begin{equation}
\begin{aligned}
\vec{c}_1&=(0,0,1)^\rmT,\\ \vec{c}_2&=\frac{1}{2}(\sqrt{3},0,-1)^\rmT,\\ \vec{c}_3&=\frac{1}{2}(-\sqrt{3},0,-1)^\rmT.
\end{aligned}
\end{equation}
The resulting protocol has the least complexity cost of 12 that is inevitable \cite{SaunPPS12}.
Therefore, to demonstrate one-way steering, it suffices to employ the simplest setting that is able to demonstrate steering.

\section{Summary}
\label{sec:summary}
In summary, we studied  steering systematically in the simplest nontrivial scenario.
We showed that  a two-qubit state can generate  EPR-nonlocal full correlations  in this scenario iff it can generate Bell-nonlocal correlations. When full statistics is taken into account,  surprisingly, the same  scenario can demonstrate one-way steering and  the hierarchy between steering and Bell nonlocality in the simplest and strongest form. To illustrate these intriguing phenomena, we provided several fiducial  examples, which can be realized in real experiments with current technology. In the course of study, we  introduced the concept of restricted  LHS  models and thereby derived a simple SDP criterion  to determine the steerability of any bipartite state under given measurements. Analytical criteria  are further derived in several important scenarios.

Our work prompts several interesting questions, which deserve further study. For example, what is the explicit analytical necessary and sufficient criterion on steerability in the simplest steering scenario (two dichotomic measurements for both Alice and Bob)? Does there exist a two-qubit state that is not steerable in the simplest scenario,
but is steerable in the second simplest scenario in which Alice performs two dichotomic
measurements and Bob performs full tomography (by either three
dichotomic measurements or a single four-outcome informationally
complete POVM)? We hope that these questions will stimulate further progress in the study of steering.

Note added: Recently, we noticed that our  work had some overlap with Ref.~\cite{CostA15}.

\section*{Acknowledgment}
We are grateful to one referee for suggesting an alternative proof of \thref{thm:BellSteer}.
Q.Q  and W.L.Y. acknowledge  supports  from The National Natural Science Foundation of China (NSFC), Grants No.
 11375141, No. 11425522, and No. 11434013. H.Z.   acknowledges financial supports
from the Excellence Initiative of the German Federal and State Governments
(ZUK81) and the DFG as well as the Perimeter Institute for
Theoretical Physics. Research at the Perimeter Institute is
supported by the Government of Canada through Industry Canada and by the Province of Ontario through the
Ministry of Research and Innovation. H.F. is supported by Ministry of Science and Technology of China (Grants No. 2016YFA0302104 and No. 2016YFA0300600), NSFC (Grant No. 91536108), and CAS (Grants No. XDB01010000 and No. XDB21030300).

\appendix

\section{\label{asec:Hierarchy}The hierarchy  between entanglement, EPR steering,  and Bell nonlocality}
In this appendix, we briefly review the hierarchy between entanglement, EPR steering,  and Bell nonlocality for the convenience of the reader. The concepts of entanglement, EPR steering, and Bell nonlocality all originated from the EPR paradox \cite{EinsPR35,Schr35}. For a long time, the three concepts were thought to be equivalent to each other  since researchers  only considered pure states. Starting with the work of Werner  in 1989 \cite{Wern89}, researchers gradually realized  that the three concepts are, in general, very different for mixed states. There exists a logical hierarchy between them: Bell nonlocality is stronger than steering, while steering is stronger than entanglement.

In 2007, Wiseman \emph{et al.} clarified the operational as well as  mathematical distinctions between   entanglement, EPR steering, and Bell nonlocality \cite{WiseJD07, JoneWD07}. In addition, they proved that, when restricted  to projective measurements,  steerability is strictly stronger than nonseparability, and strictly weaker than Bell nonlocality.  In other words, every violation of a Bell inequality   demonstrates  EPR steering, and every violation of  an EPR-steering  inequality demonstrates entanglement (but the converses do not hold in general). Two-qubit Werner states provide a concrete demonstration of this hierarchy.  Furthermore,  recently the strict  hierarchy was shown to hold even under generalized measurements, as represented by positive-operator-valued measures (POVMs)  \cite{QuinVC15}.

The strict hierarchy between entanglement, EPR steering,  and Bell nonlocality is also reflected in the complexity costs for demonstrating them experimentally, as pointed out by Saunders \emph{et al}. \cite{SaunPPS12}. Here the complexity cost is defined as
\begin{equation}
W=\prod_{p=1}^P\sum_{s=1}^{S_p}O_p^s,
\end{equation}
where $P\geq2 $  represents the number of  distinct parties in the experimental demonstration,  $S_p\geq1 $ represents the number of different measurement settings employed by party~$p$,  $O_p^s\geq2 $ represents the number of potential measurement outcomes for party $p$ when setting $s$ is chosen,  and $W$ represents the total number of potential joint measurement outcomes that may appear. The authors of  \rcite{SaunPPS12} showed that  the least complexity costs for demonstrating entanglement, EPR steering,  and Bell nonlocality are  $W_\rmE=9 $,  $W_\rmS=12$, and  $W_\rmB=16 $, respectively.

In certain special cases, the hierarchy mentioned above may not always be strict. In the case of  pure states, for example, it is well known that entanglement is both necessary and sufficient for demonstrating steering and Bell nonlocality. If the steering party can only perform two projective measurements, then a
two-qubit Bell-diagonal state can demonstrate steering iff it can demonstrate Bell nonlocality~\cite{QuanZLF16}. In addition, the least complexity costs for demonstrating entanglement, EPR steering,  and Bell nonlocality under projective measurements are  all equal to 16.
In general, it is highly  nontrivial to determine the precise relations between  entanglement, EPR steering,  and Bell nonlocality for a given scenario.

All  three concepts, i.e., entanglement, EPR steering, and Bell nonlocality, are useful  in certain contexts.  For example, all of them are  useful to QKD. To be specific, Bell nonlocality is necessary for realizing device-independent
QKD, while EPR steering is necessary for realizing one-sided device-independent QKD \cite{TomaR11,BranCWS12}. In practice, it is usually much more difficult to demonstrate Bell nonlocality  than EPR steering. Therefore,  understanding the hierarchy between the three forms of nonlocality is of fundamental interest not only to foundational studies but also to practical applications.
Our work is a contribution along this direction.

\section{\label{asec:CSBProofAlt}Alternative proof of Theorem~2}
In this appendix, we provide an alternative proof of Theorem~2 in the main text, which offers a complementary perspective  on the relation between steering and Bell nonlocality.

To start with, we clarify the relations between steering and Bell nonlocality for Bell-diagonal states in the simplest scenario based on an earlier work \cite{QuanZLF16}.
\begin{lemma}\label{lem:BellEPR}
	The following five statements concerning a Bell-diagonal state $\rho$ are equivalent.
	\begin{enumerate}
		\item $\rho$ is steerable by two projective measurements.
		
		\item $\rho$ is steerable  in the simplest steering scenario.
		
		\item $\rho$ can generate EPR-nonlocal full correlations in the simplest steering scenario.
		
		\item $\rho$ is Bell nonlocal in the simplest Bell scenario.
		
		\item $\rho$ can generate Bell-nonlocal full correlations in the simplest Bell scenario.
	\end{enumerate}
\end{lemma}
\begin{remark}
	Here the simplest steering scenario and Bell scenario mean two projective measurements for each party.
	Statements 4 and 5 are equivalent since the CHSH inequality, a full-correlation inequality, is the only Bell inequality in the simplest Bell scenario \cite{Scar13}.
	 However, we are not aware of any simple and intuitive proof of this latter fact. 	For Bell-diagonal states, the equivalence also follows from the fact that the marginal statistics are completely random.
	 Remarkably, the following proof of the lemma does not rely on  this equivalence.
\end{remark}

\begin{proof}
	EPR-nonlocal full correlations imply steerability, and Bell-nonlocal full correlations imply Bell nonlocality.
	Thanks to the hierarchy between steering and Bell nonlocality,	
	any of statements~2 to~5 implies statement~1; 	statement~5 implies  statements~1 to~4. According to our early work in \rcite{QuanZLF16}, a Bell-diagonal state is steerable by two projective measurements iff it can violate the CHSH inequality, that is, iff it can generate Bell-nonlocal full correlations in the simplest Bell scenario. Therefore, statements~1 and~5 are equivalent, which implies the equivalence of all five statements in view of the above observation.
\end{proof}

\begin{proof}[Alternative proof of Theorem~2]Let $\rho$ be a two-qubit state and $\varrho$ the Bell-diagonal state with the same correlation matrix as $\rho$. Then, $\varrho$ and $\rho$ generate the same full correlations in the simplest steering and Bell scenarios. In particular, $\rho$ can generate EPR-nonlocal  full correlations iff $\varrho$ can, which is the case iff $\varrho$ can generate Bell-nonlocal (full) correlations	according to \lref{lem:BellEPR}. Therefore, $\rho$ can generate EPR-nonlocal full correlations in the simplest steering scenario iff it can generate Bell-nonlocal (full) correlations in the simplest Bell scenario.
\end{proof}
\begin{remark}
	The alternative proof   does  not yield  a criterion on generating EPR-nonlocal full correlations directly. Interestingly, the theorem is established based on the hierarchy between steering and Bell nonlocality as well as the known criterion on the violation of the CHSH inequality.
	In sharp contrast, the proof in the main text establishes  the relation between steering and  Bell nonlocality by first deriving
	the maximal violation of the analog CHSH inequality.  It  yields
	a necessary and sufficient criterion on generating EPR-nonlocal full correlations directly and also provides a natural measure of the correlation strength. Both approaches are instructive and interesting in their own rights; together
	they offer complementary perspectives  on steering and Bell nonlocality in the simplest scenario.
\end{remark}

\bibliography{Quanref}%

\end{document}